\documentclass[a4paper, 14pt, conference,onecolumn]{ieeeconf}      

\IEEEoverridecommandlockouts                              

\usepackage{graphics} 
\usepackage{epsfig} 

\usepackage{amsmath} 
\usepackage{amssymb}  
\usepackage{tabularx}
\usepackage{amsmath}
\usepackage{balance}

\newcounter{subassumption}[asu]

\makeatletter
\renewcommand{\p@subassumption}{\theasu}
\makeatother

\usepackage{rotating,booktabs,multirow}
\makeatletter
\let\NAT@parse\undefined
\makeatother
\usepackage{hyperref}
\usepackage{graphicx}
\newtheorem{remark}{Remark}[section]

\usepackage[latin1]{inputenc}
\usepackage{color}
\usepackage{epsfig}
\usepackage{graphicx}
\usepackage[hang]{caption2}
\usepackage[normalsize,it,hang]{subfigure}
\DeclareGraphicsExtensions{.jpg,.png,.jpg,.jpg}

\title{{\tt \small IFAC Symposium on Biological and Medical Systems - 11th BMS 2021 \\ Ghent, Belgium - September 19-22, 2021} \\ \ \\ \LARGE  \bf
Toward simple \textit{in silico} experiments \\ for drugs administration \\ in some cancer treatments 
}

\author{Michel Fliess$^{1, 5}$, C\'{e}dric Join$^{2, 5}$, Kaouther Moussa$^{3}$, Seddik M. Djouadi$^{4}$, Mohamed W. Alsager$^{4}$
\thanks{$^{1}$LIX (CNRS, UMR 7161), \'Ecole polytechnique, 91128 Palaiseau, France.\newline {\tt \small Michel.Fliess@polytechnique.edu } } 
\thanks{$^2$CRAN (CNRS, UMR 7039)), Universit\'{e} de Lorraine, BP 239, 54506 Vand{\oe}uvre-l\`{e}s-Nancy, France. \newline
{\tt\small Cedric.Join@univ-lorraine.fr}}
\thanks{$^3$ Universit\'{e} Grenoble Alpes, CNRS, Grenoble INP, GIPSA-lab, 38000 Grenoble, France. \newline
{\tt\small kaouther.moussa@gipsa-lab.fr}}
\thanks{$^4$ Department of Electrical Engineering and Computer Science, University of Tennessee, Knoxville, TN 37996, USA. \newline
{\tt\small mdjouadi@utk.edu, malsager@vols.utk.edu}}
\thanks{$^{5}$AL.I.E.N. (ALg\`ebre pour Identification \& Estimation Num\'eriques), 7 rue Maurice Barr\`{e}s, 54330 V\'{e}zelise, France. \newline
        {\tt \small \{cedric.join, michel.fliess\}@alien-sas.com}}
        }

\begin{document}
\maketitle
\thispagestyle{empty}
\pagestyle{empty}

\begin{abstract}
We present some ``in silico'' experiments to design  combined chemo- and immunotherapy  treatment schedules. We introduce a new framework by combining flatness-based control, which is a model-based setting, along with model-free control. The flatness property of the used mathematical model yields straightforward reference trajectories. They provide us with the nominal open-loop control inputs. Closing the loop via model-free control allows to deal with the uncertainties on the injected drug doses. Several numerical simulations illustrating different case studies are displayed. We show in particular that the considered health indicators are driven to the safe region, even for critical initial conditions. Furthermore, in some specific cases there is no need to inject chemotherapeutic agents.
\end{abstract}

\newpage

\section{Introduction}
We consider drug injections scheduling for cancer treatments from a control point of view (see, \textit{e.g.}, Chap. 10 in \cite{qatar} for bibliographical references). Among the many models which have been used, those stemming from an earlier work of \cite{stepanova} are quite popular.  Most appealing are several publications by d'Onofrio and different coauthors: see especially \cite{ono12}. Such approaches to chemo- and immunotherapy led in recent years to promising control-theoretic investigations: see, \textit{e.g.}, \cite{alamir14}; \cite{scha} and references therein; \cite{moussa,sharifi1,sharifi2}. They employ various optimization techniques which are related to optimal control, model predictive control, and robust control.

We explore here another route via tools which are combined here for the first time, although they both gave rise to an abundant literature in control engineering:
\begin{enumerate}
\item \emph{Flatness}-based control (see \cite{ijc,ieee}; and \cite{sira,levine,rudolph}) is a model-based approach which has been well received in many industrial domains. See, \textit{e.g.}, \cite{bonnabel} for tower cranes.
\item Besides being useful in concrete case-studies (see, \textit{e.g.}, \cite{oak,parkol,park,telsang,tumin} for energy management), \emph{model-free} control in the sense of \cite{mfc13,nicu} has already been illustrated in biomedicine (\cite{nantes,bara,faraji}) and in bioengineering (\cite{ventil}). Note that the terminology ``model-free control'' has been used many times with different definitions: see \cite{mfcalamir} in oncology. 
\end{enumerate}
Our \emph{virtual patient} is modeled through two ordinary differential equations presented by \cite{ono12}. This system is trivially flat with obvious \emph{flat outputs}. The design of suitable reference trajectories with the corresponding open-loop controls becomes straightforward. A major source of uncertainty, according to \cite{sharifi2}, is the unknown fluctuation of the drug delivery to the tumor, which should be related to actuators faults, \textit{i.e.}, to a classic topic in fault-tolerant control (see, \textit{e.g.}, \cite{fault}). It has been already noticed that model-free control is well-suited for dealing with actuators faults: see \cite{mfc13} for an academic example and \cite{toulon} for a concrete case-study. The loop is therefore closed via model-free control. Let us emphasize the following points:
\begin{itemize}
\item The computer implementation is easy.
\item Only a low computing cost is necessary. 
\item Some \emph{scenarios}, \textit{i.e.}, \emph{in silico} experiments, lead to unexpected results. They might attract cancerologists. 
\end{itemize}

Our paper is organized as follows. Section~\ref{Sec_model} { presents the dynamical model of the virtual patient}, Section \ref{modelcontrol} reviews briefly  flatness-based control, and model-free control. Numerical simulations are presented in Section \ref{silico}. Section \ref{conclusion} contains some suggestions for for future research on:  1) the possible medical impact of our in silico experiments, 2) some aspects related to \emph{systems biology}, 3) new control paradigmes which might be derived from the methods developed here.

See \cite{biorxiv} for a first draft.


\section{Virtual patient  dynamical model}\label{Sec_model}

We consider the model presented in \cite{ono12}
\begin{eqnarray}
\label{1}
\dot x &= -\mu_C x\ln\left(\frac{x}{x_\infty}\right)-\gamma xy - xu\eta_x \\
\label{2}
\dot y &= \mu_I\left(x-\beta x^2\right)y-\delta y+\alpha+ yv\eta_y
\end{eqnarray}
$x$, $y$ are, respectively, the number of tumor cells and the immune cell density;  the control variables $u$ and $v$ are the cytotoxic and immune-stimulation drugs; the parameters $\mu_C$, $\mu_I$, $\alpha$, $\gamma$, $\delta$, $x_\infty$ are positive. The terms $\eta_x$, $\eta_y$, $0 \leq \eta_x \leq 1$, $0 \leq \eta_y \leq 1$, are inspired by  \cite{sharifi2}: they represent the uncertain and fluctuating parts of drugs which are delivered to the tumor. The definition as well as the numerical values of these parameters can be found in Table~\ref{Table2D}.

\begin{table}[!h]
\begin{center}
\scalebox{0.87}{
\begin{tabular}{|c|c|c|}
\hline
Parameter     & Definition                           & Numerical Value \\ \hline
${\mu_{C}}$   & tumor growth rate                    & $1.0078\cdot10^{7}$ cells/day \\
${\mu_{I}}$   & tumor stimulated proliferation rate  & .0029 ${day^{-1}}$ \\
${\alpha}$    & rate of immune cells influx          & .0827 ${day^{-1}}$  \\
${\beta}$     & inverse threshold                    & .00.31   \\
${\gamma}$    & interaction rate                     & $1\cdot10^{7}$ cells/day  \\
${\delta}$      & death rate                         & .1873 ${day^{-1}}$  \\
${\eta_X}$    & chemotherapeutic killing parameter &  $1\cdot10^{7}$ cells/day  \\
${\eta_Y}$    & immunotherapy injection parameter  & $1\cdot10^{7}$ cells/day   \\
${x_{\infty}}$ & fixed carrying capacity            &  $780 \cdot10^{6}$ cells  \\
\hline
\end{tabular}
}
\end{center}
\caption{}
\label{Table2D}
\end{table}

This system has three equilibria corresponding to $\dot{x} = \dot{y} = u = v =0$: 
\begin{enumerate}
\item a locally stable equilibrium $x = 73$, $y = 1.32$ which corresponds to a benign case; 
\item an unstable saddle point $x = 356.2$, $y = 0.439$, which separates the benign and malignant regions; 
\item a locally stable equilibrium $x = 737.3$, $y = 0.032$, which is malignant. 
\end{enumerate}

The idea behind controlling such systems consists in driving the state trajectories from the region of attraction of the malignant equilibrium (critical case) to the region of attraction of the benign equilibrium. The simulations that are presented in this paper will show that the state trajectories are driven to the benign equilibrium under control action (drug delivery) for different settings.

\section{Control methodology}\label{modelcontrol}

\subsection{Flatness property}
A control system with $m$ independent control variables is said to be \emph{(differentially) flat} if, and only if, there exists $m$ system variables $y_1, \dots, y_m$, the \emph{flat outputs}, such that any system variable $z$, the control variables for instance, may be expressed as a \emph{differential} function of $y_1, \dots, y_m$, \textit{i.e.}, $z = \Phi (y_1, \dots, y_m, \dots, y_1^{(\nu_1)}, \dots, y_m^{(\nu_m)})$, where the derivation orders $\nu_1, \dots, \nu_m$ are finite. A linear system is flat if, and only if, it is controllable. Thus flatness may be viewed as another extension of Kalman's controllability.

Equations \eqref{1}-\eqref{2} yield
\begin{eqnarray*}
\label{3}
u &= \frac{\dot x  + \mu_C x\ln\left(\frac{x}{x_\infty}\right) + \gamma xy}{- x\eta_x} = X(x, \dot{x}, y) \\
\label{4}
v &= \frac{\dot y - \mu_I\left(x-\beta x^2\right)y + \delta y - \alpha}{y\eta_y} = Y(y, \dot{y}, x)
\end{eqnarray*}
The above equations show immediately that System  \eqref{1}-\eqref{2} is flat; $x$, $y$ are flat outputs.

\subsection{Reference trajectory and nominal open-loop control}
One of the main benefits of flatness is the possibility of easily deriving a suitable reference trajectory and the corresponding nominal open-loop control. For a given reference trajectory $x^\star(t)$, $y^\star(t)$, the corresponding nominal control variables 
\begin{eqnarray}
\label{shoot1}
u^\star (t) &= X(x^\star (t), \dot{x}^\star (t), y^\star (t)) \\
\label{shoot2}
v^\star (t) &= Y(y^\star (t), \dot{y}^\star (t), x^\star (t))
\end{eqnarray}

might exhibit unacceptable negative values. Define therefore the nominal open-loop control variables 
\begin{eqnarray*}
u_{\rm OL}(t) &= u^\star (t) \, {\rm if} \, u^\star (t) \geq 0, \; u_{\rm OL}(t) = 0 \, {\rm if} \, u^\star (t) < 0  \\
v_{\rm OL}(t) &= v^\star (t) \, {\rm if} \, v^\star (t) \geq 0, \; v_{\rm OL}(t) = 0 \, {\rm if} \, v^\star (t) <  0
\end{eqnarray*}

\subsection{Closing the loop via model-free control}\label{loop}
From a control-engineering standpoint the terms $\eta_x$ and $\eta_y$ should be related to actuators faults. Introduce therefore the two ``decoupled'' \emph{ultra-local models} (\cite{mfc13,toulon}):
$$
\dot{z}_x = F_x + \alpha_x u_{\rm MFC}, \ \dot{z}_y = F_y + \alpha_y v_{\rm MFC} 
$$
where $z_x = x - x^\star$, $z_y = y - y^\star$ are the tracking errors; $\alpha_x$ (resp. $\alpha_y$) is a constant parameter which is chosen by the practitioner such that $\dot{x}$ and $\alpha_x u$ (resp. $\dot{y}$ and $\alpha_y v$) are of the same order of magnitude; $F_x$ and $F_y$, which are data-driven, subsume the poorly known structures and disturbances. A real-time estimation (\cite{mfc13}) of $F_x$, $F_y$ are given by {\small $$F_x^{\rm est} = - \frac{6}{\tau_x^{3}} \int_{t-\tau_x}^t \left( (t-2\sigma)x(\sigma) + \alpha_x \sigma(\tau_x - \sigma)u_{\rm MFC}(\sigma)\right)d\sigma$$} {\small $$F_y^{\rm est} = - \frac{6}{\tau_y^{3}} \int_{t-\tau_y}^t \left( (t-2\sigma)y(\sigma) + \alpha_y \sigma(\tau_y - \sigma)v_{\rm MFC}(\sigma)\right)d\sigma$$} 

\noindent{where} $\tau_x, \tau_y > 0$ are ``small.'' Close the loop via an \emph{intelligent Proportional} controller, or \emph{iP}, 
$$u_{\rm MFC} = - \frac{F_x^{\rm est} + K_{x, P} z_x}{\alpha_x}, \ v_{\rm MFC} = - \frac{F_y^{\rm est} + K_{y, P} z_y}{\alpha_y}$$ where $K_{x, P}, K_{y, P} > 0$. From $\dot{z}_x + K_{x, P} z_x = 0$, $\dot{z}_y + K_{y, P} z_y = 0$, it follows that those two gains ensure local stability around the reference trajectory.

The close-loop controls $u_{\rm CL}$, $v_{\rm CL}$ may now be defined:
\begin{itemize}
\item If $u_{\rm OL} + u_{\rm MFC} \geq 0$, then $u_{CL} = u_{\rm OL} + u_{\rm MFC}$; if $u_{\rm OL} + u_{\rm MFC} < 0$, then $u_{\rm CL} = 0$. 
 \item If $v_{\rm OL} + v_{\rm MFC} \geq 0$, then $v_{CL} = v_{\rm OL} + v_{\rm MFC}$; if $v_{\rm OL} + v_{\rm MFC} < 0$, then $v_{\rm CL} = 0$.
\end{itemize}

\section{Numerical simulations}\label{silico}
\subsection{Presentation}\label{prez}
\subsubsection{A shooting method}
A huge number in silico experiments have been most easily performed via the flatness property, \textit{i.e.}, via Formulae \eqref{shoot1}-\eqref{shoot2}. It permits to select the most suitable ones with respect to boundary conditions, optimality criteria and constraints. Our approach might appear therefore as an alternative to the \emph{shooting} methods in optimal control and numerical analysis (see, \textit{e.g.}, \cite{shooting,pel}).

\subsubsection{Time sampling and duration}
The duration of an experiment is $60$ days. The time sampling interval is equal to $30$ minutes. 
\begin{remark}
The total simulations duration is 60 days, even though the figures are limited to 30 days for visibility reasons, since all the variables reach a steady state at this time.
\end{remark}

\subsection{Closed-loop and total amount of drugs}\label{nominal}
\begin{figure*}[!ht]
\centering%
\subfigure[\footnotesize Control $u$ (blue $--$) and Nominal control $u^\ast$ (black $- -$) ]
{\includegraphics[width=0.45\textwidth]{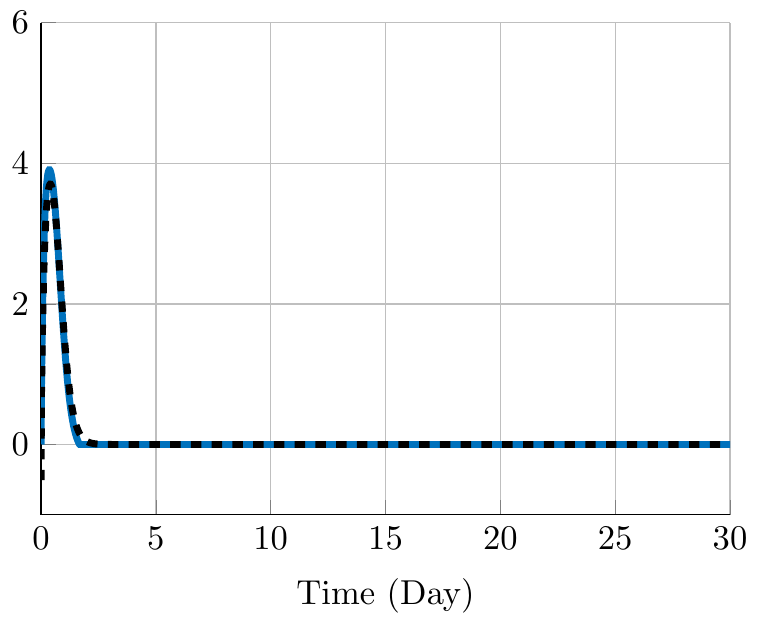}}
\subfigure[\footnotesize Control $v$ (blue $--$) and Nominal control $v^\ast$ (black $- -$)]
{\includegraphics[width=0.45\textwidth]{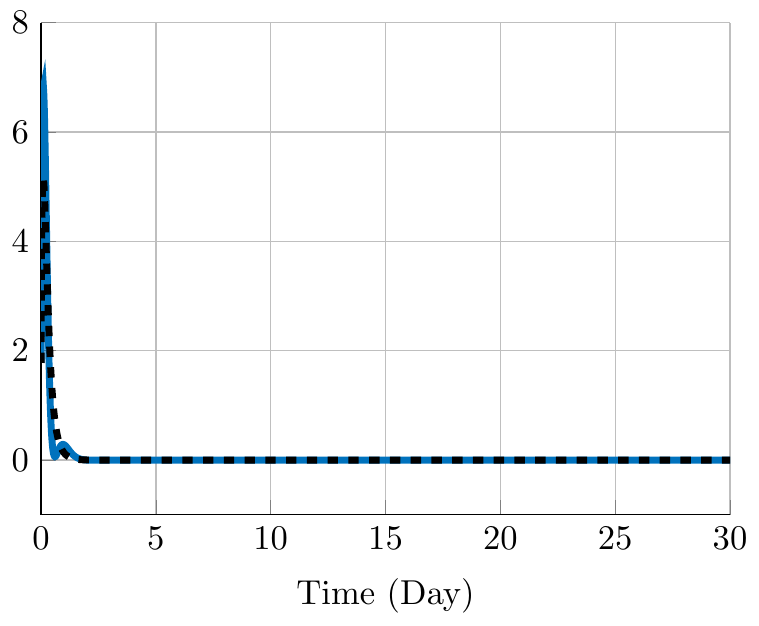}}
\\
\subfigure[\footnotesize Output $x$ (blue $--$), Reference trajectories (black $- -$) and Stable points (red and green $-.$)]
{\includegraphics[width=0.45\textwidth]{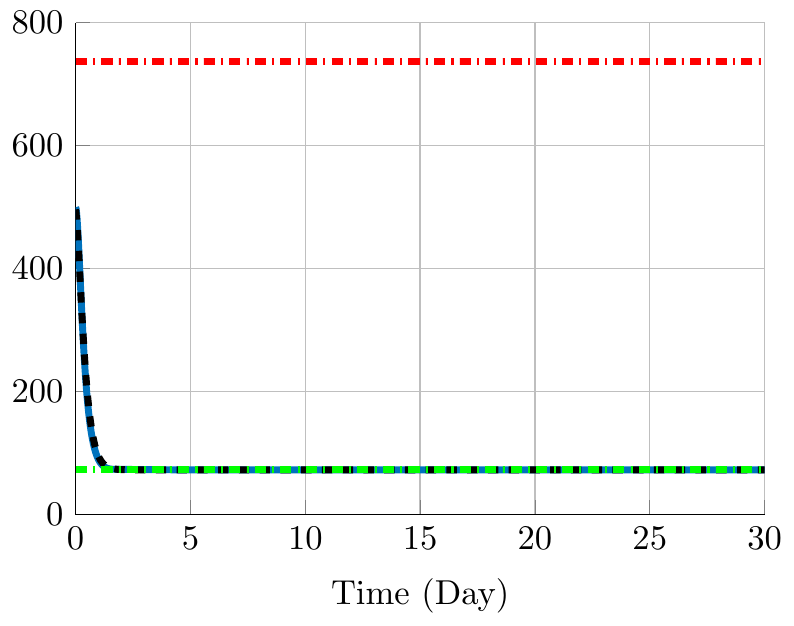}}
\subfigure[\footnotesize Output $y$ (blue $--$), Reference trajectories (black $- -$) and Stable points (red and green $-.$)]
{\includegraphics[width=0.45\textwidth]{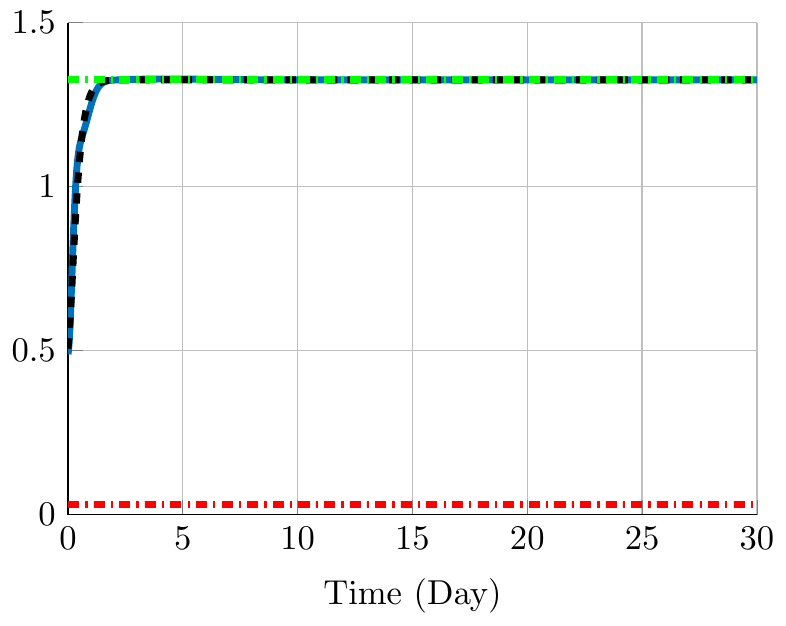}}
\caption{Fast trajectory}\label{S1}
\end{figure*}

\begin{figure*}[!ht]
\centering%
\subfigure[\footnotesize Control $u$ (blue $-$) and Nominal control $u^\ast$ (black $- -$) ]
{\includegraphics[width=0.45\textwidth]{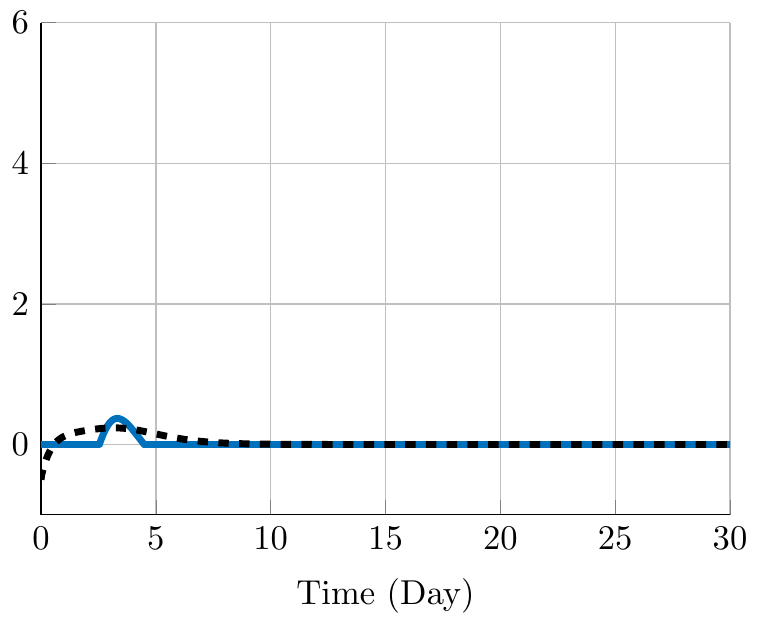}}
\subfigure[\footnotesize Control $v$ (blue $-$) and Nominal control $v^\ast$ (black $- -$)]
{\includegraphics[width=0.45\textwidth]{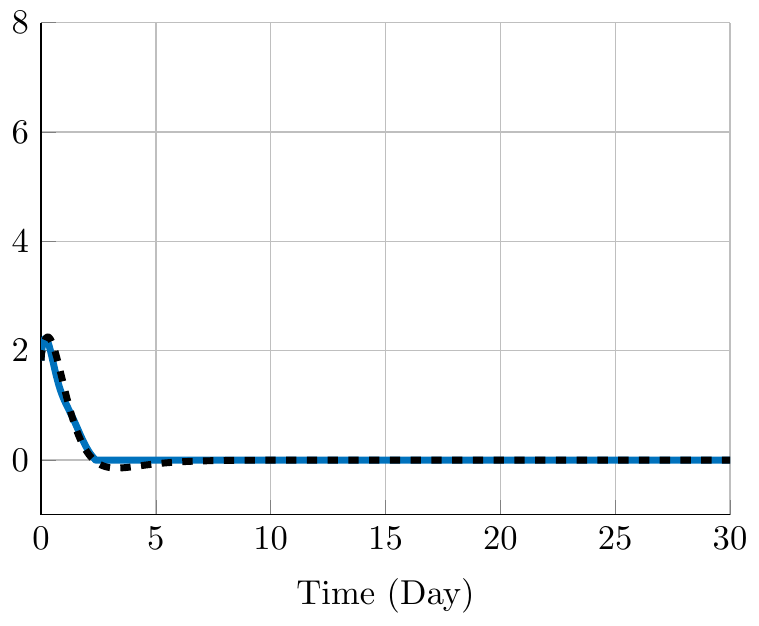}}
\\
\subfigure[\footnotesize Output $x$ (blue $-$), Reference trajectories (black $- -$) and Stable points (red and green $-.$)]
{\includegraphics[width=0.45\textwidth]{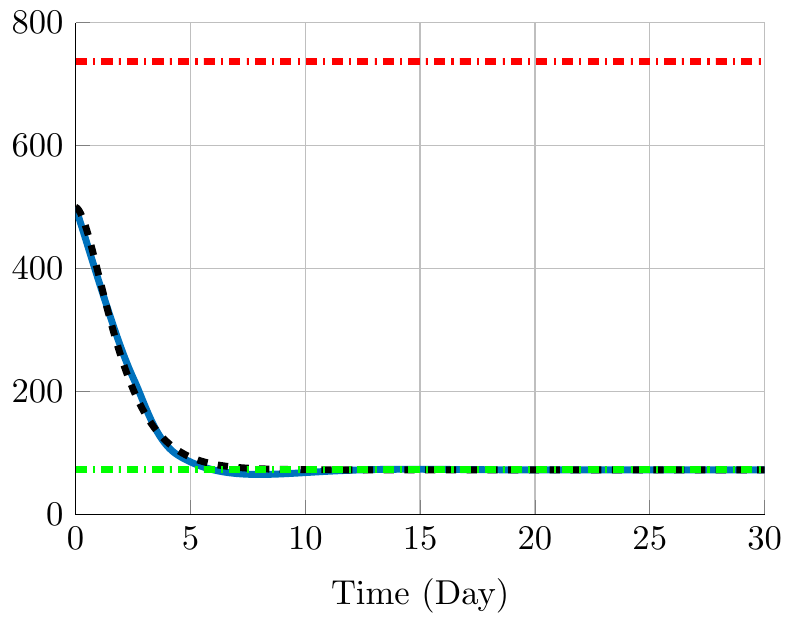}}
\subfigure[\footnotesize Output $y$ (blue $-$), Reference trajectories (black $- -$) and Stable points (red and green $-.$)]
{\includegraphics[width=0.45\textwidth]{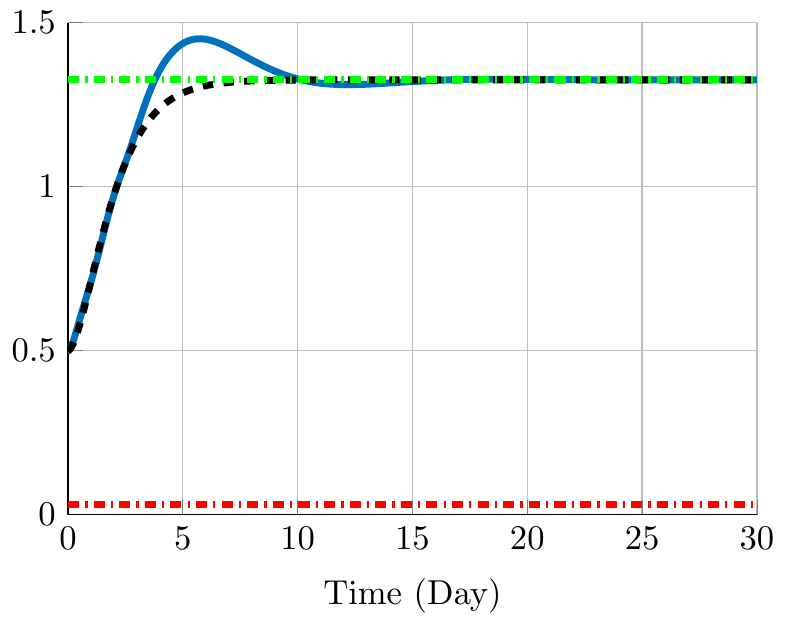}}
\caption{Slow trajectory}\label{S2}
\end{figure*}

\begin{figure*}[!ht]
\centering%
\subfigure[\footnotesize Control $u$ integral of figure \ref{S1}-(a)]
{\includegraphics[width=0.45\textwidth]{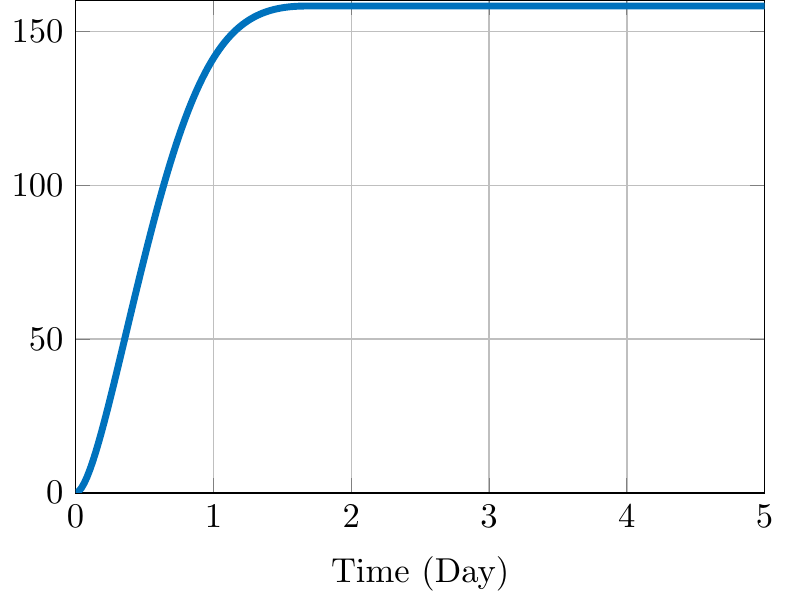}}
\subfigure[\footnotesize Control $v$ integral of figure \ref{S1}-(b)]
{\includegraphics[width=0.45\textwidth]{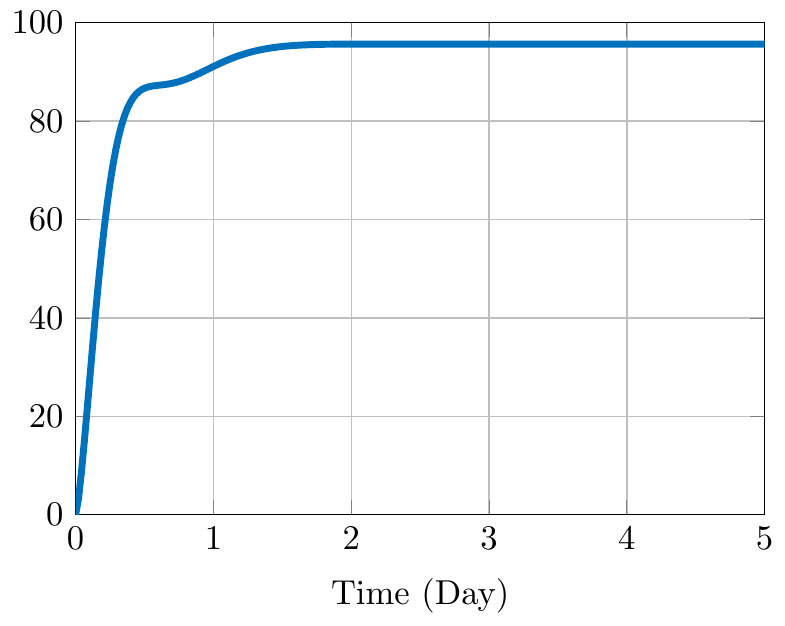}}
\\
\subfigure[\footnotesize Control $u$ integral of figure \ref{S2}-(a)]
{\includegraphics[width=0.45\textwidth]{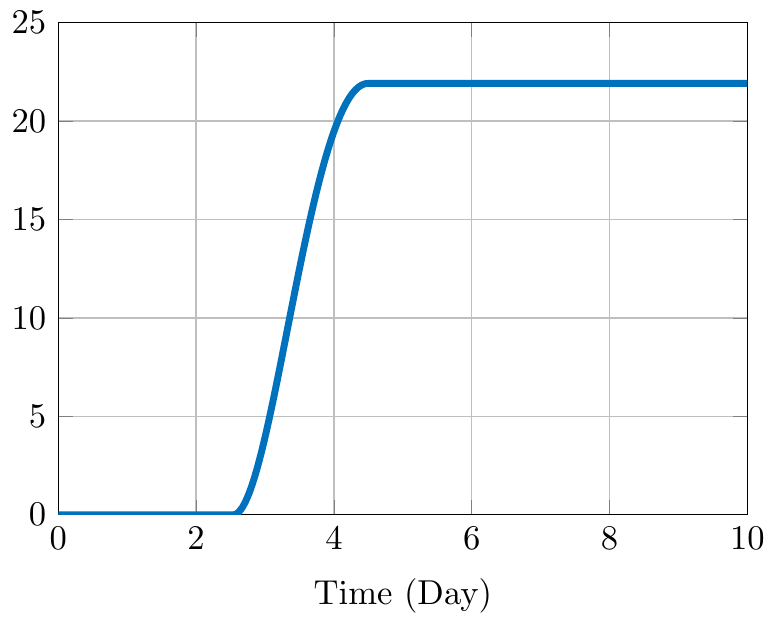}}
\subfigure[\footnotesize Control $v$ integral of figure \ref{S2}-(b)]
{\includegraphics[width=0.45\textwidth]{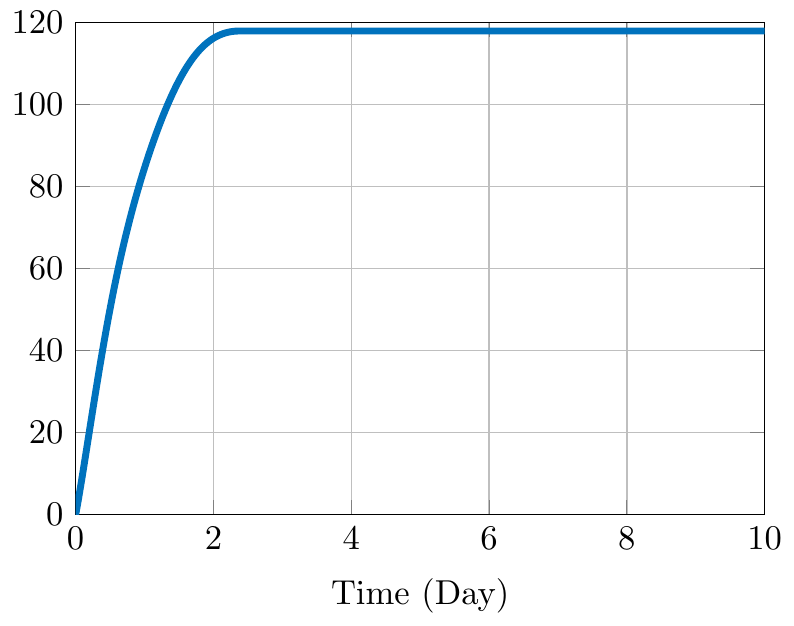}}
\caption{Comparison between total drug injections}\label{S12}
\end{figure*}

Set $\eta_x = \eta_y = 0.5$. This nominal value might be large according to \cite{sharifi2}.  Figures \ref{S1} and \ref{S2} display two experiments with the same initial point $x = 500$, $y = 0.5$, which lies in the attraction region of the malignant equilibrium. The total amounts of injected drugs, which are often considered as important constraints, are given by the two integrals $\int_0^T u_{\rm CL}(\tau) d\tau$, $\int_0^T v_{\rm CL}(\tau) d\tau$, where $T$ is the experiment duration. Figure \ref{S12} indicates that the quantity of drugs injected during the slow scenario is lower than in the fast one. This outcome ought to be discussed in oncology.

\subsection{Other scenarios}
\subsubsection{Same initial point.} 
\begin{figure*}[!ht]
\centering%
\subfigure[\footnotesize Control $u$ (blue $-$) and Nominal control $u^\ast$ (black $- -$) ]
{\includegraphics[width=0.45\textwidth]{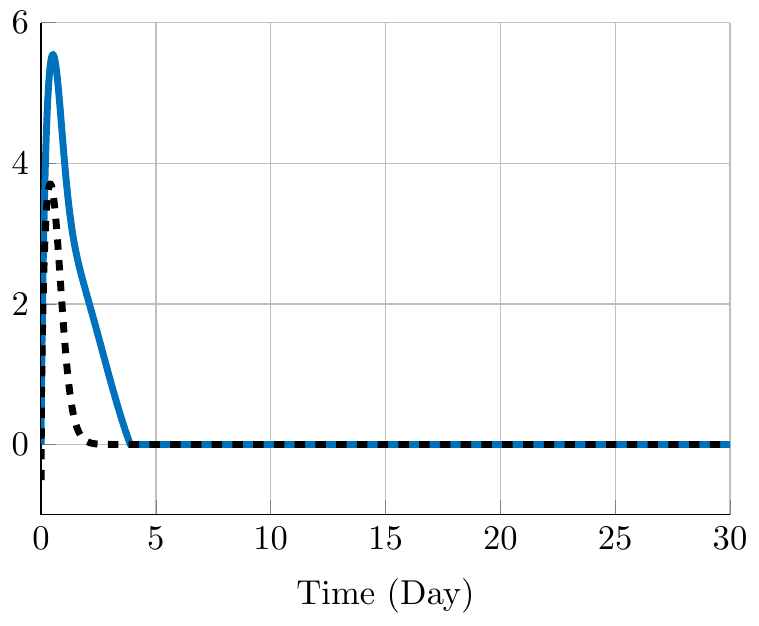}}
\subfigure[\footnotesize Control $v$ (blue $-$) and Nominal control $v^\ast$ (black $- -$)]
{\includegraphics[width=0.45\textwidth]{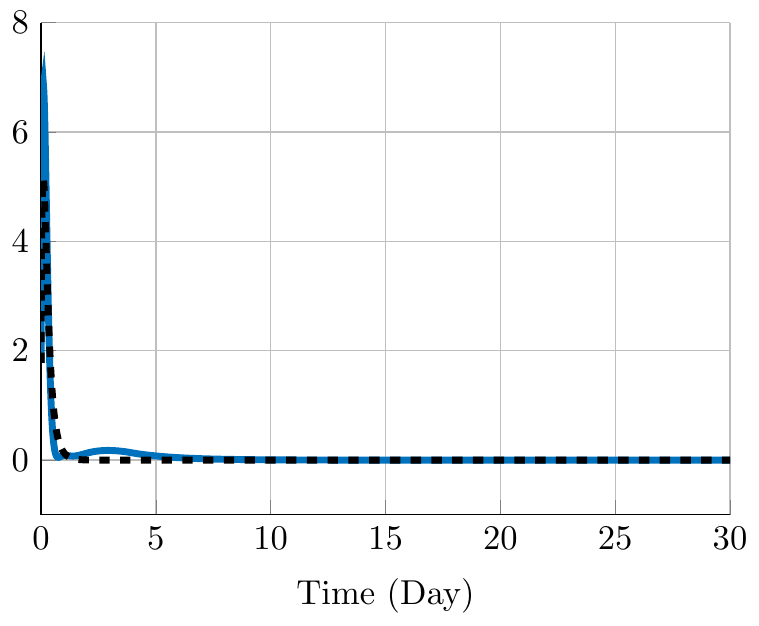}}
\\
\subfigure[\footnotesize Output $x$ (blue $-$), Reference trajectories (black $- -$) and Stable points (red and green $-.$)]
{\includegraphics[width=0.45\textwidth]{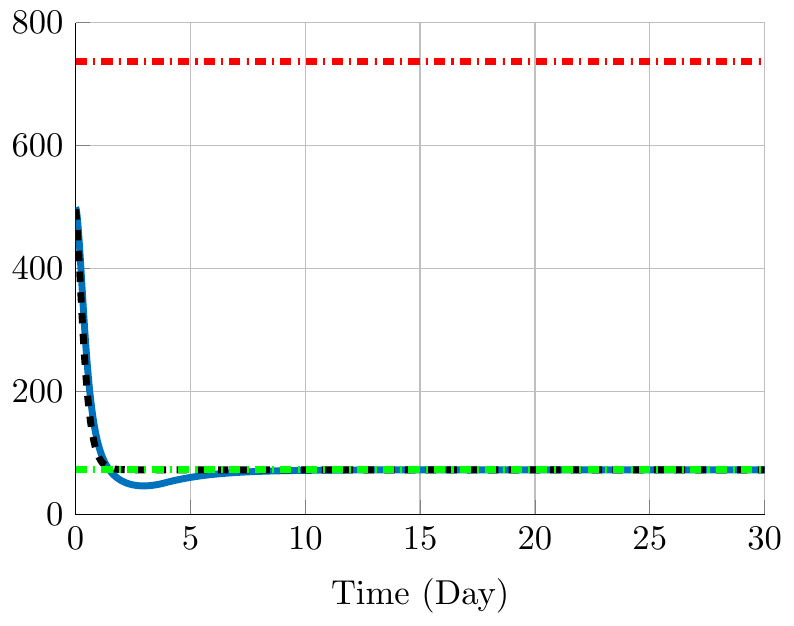}}
\subfigure[\footnotesize Output $y$ (blue $-$), Reference trajectories (black $- -$) and Stable points (red and green $-.$)]
{\includegraphics[width=0.45\textwidth]{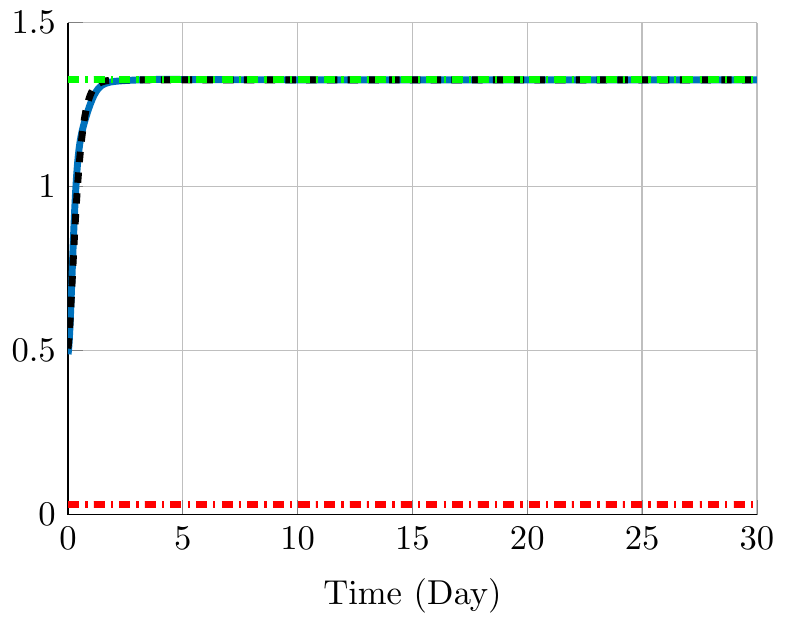}}
\caption{Unknown variation of $\eta_x$}\label{S3}
\end{figure*}

\begin{figure*}[!ht]
\centering%
\subfigure[\footnotesize Control $u$ (blue $-$) and Nominal control $u^\ast$ (black $- -$) ]
{\includegraphics[width=0.45\textwidth]{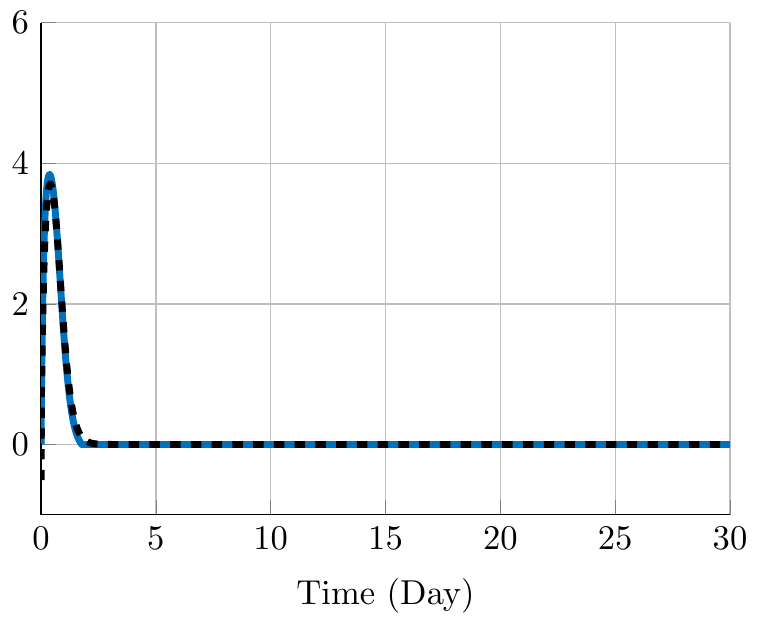}}
\subfigure[\footnotesize Control $v$ (blue $-$) and Nominal control $v^\ast$ (black $- -$)]
{\includegraphics[width=0.45\textwidth]{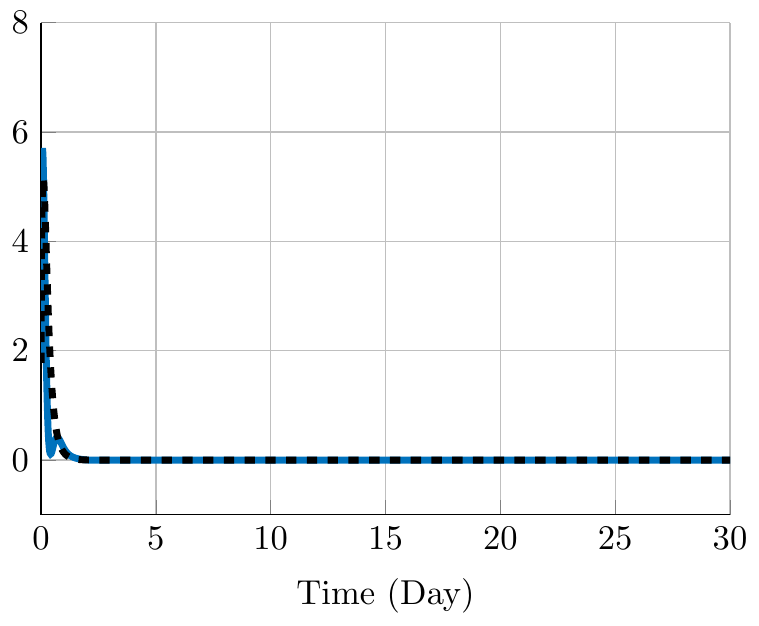}}
\\
\subfigure[\footnotesize Output $x$ (blue $-$), Reference trajectories (black $- -$) and Stable points (red and green $-.$)]
{\includegraphics[width=0.45\textwidth]{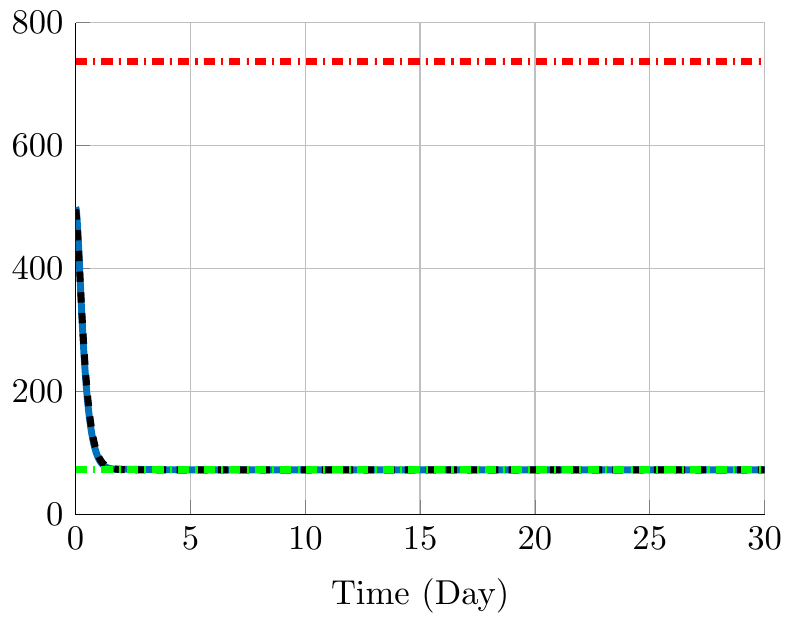}}
\subfigure[\footnotesize Output $y$ (blue $-$), Reference trajectories (black $- -$) and Stable points (red and green $-.$)]
{\includegraphics[width=0.45\textwidth]{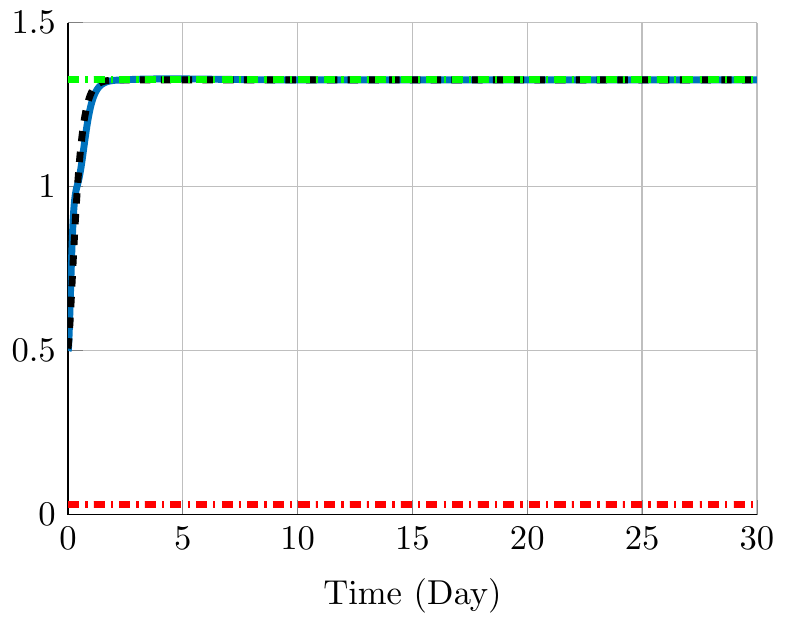}}
\caption{Unknown variation of $\eta_y$}\label{S4}
\end{figure*}

Here $\eta_x = 0.31$, $\eta_y = 0.75$ are supposed to be unknown. Use the same nominal parameters as in Section \ref{nominal}, and the feedback loop of Section \ref{loop}, with $\alpha_x = - 10000$, $\alpha_y = 1$, $K_{x, P} = 100$, $K_{y, P} = 10$. The results depicted in Figures \ref{S3} and \ref{S4} show that the benign equilibrium is reached after a short period of time.
\subsubsection{New initial point.} 
\begin{figure*}[!ht]
\centering%
\subfigure[\footnotesize Control $u$ (blue $-$) and Nominal control $u^\ast$ (black $- -$) ]
{\includegraphics[width=0.45\textwidth]{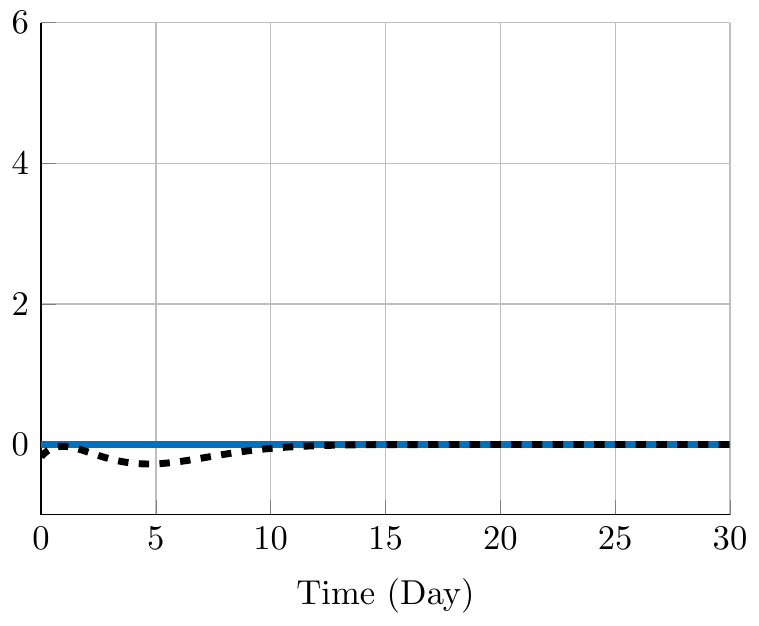}}
\subfigure[\footnotesize Control $v$ (blue $-$) and Nominal control $v^\ast$ (black $- -$)]
{\includegraphics[width=0.45\textwidth]{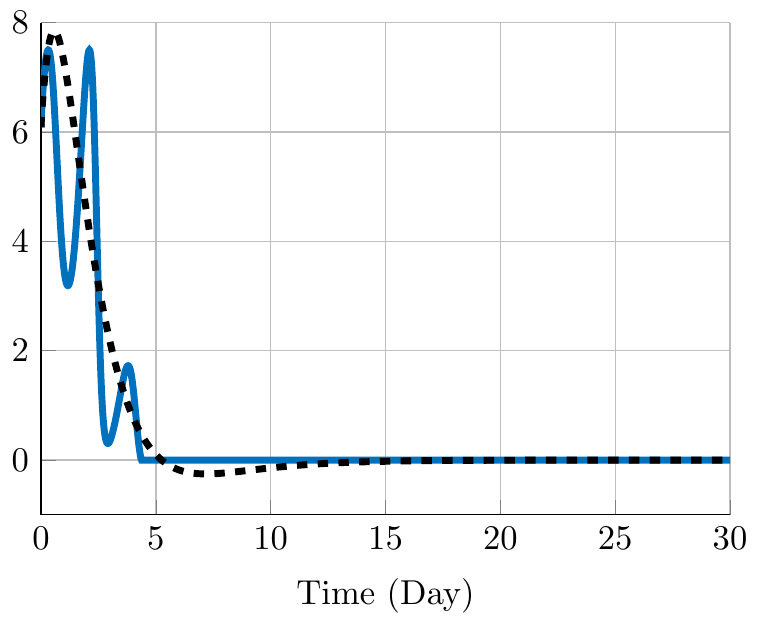}}
\\
\subfigure[\footnotesize Output $x$ (blue $-$), Reference trajectories (black $- -$) and Stable points (red and green $-.$)]
{\includegraphics[width=0.45\textwidth]{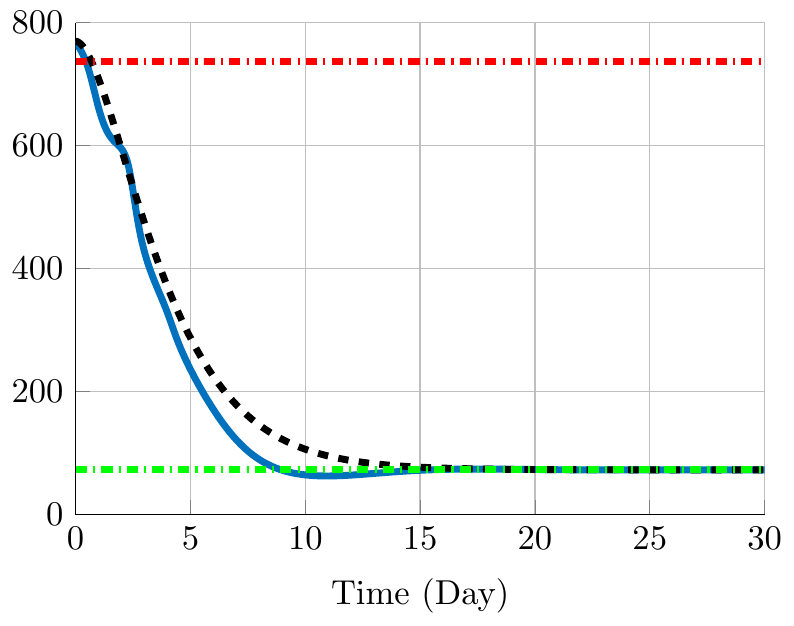}}
\subfigure[\footnotesize Output $y$ (blue $-$), Reference trajectories (black $- -$) and Stable points (red and green $-.$)]
{\includegraphics[width=0.45\textwidth]{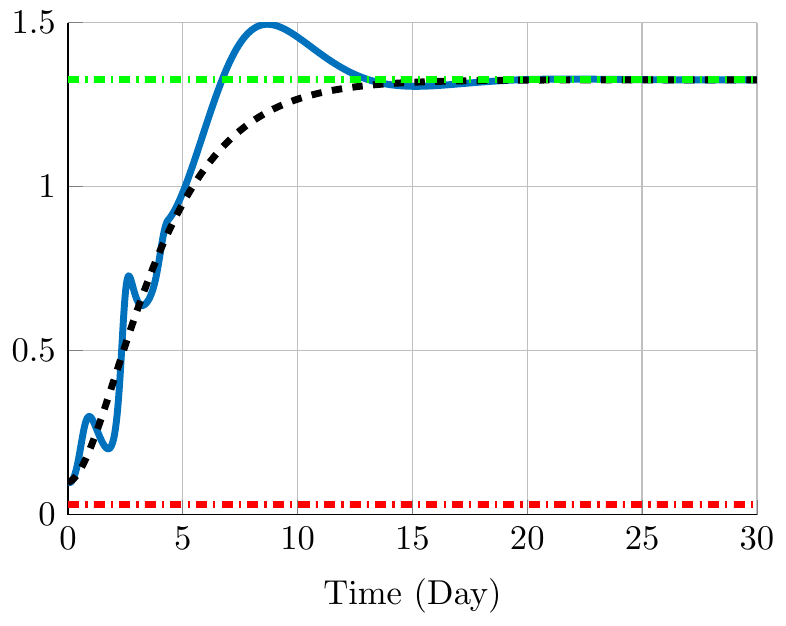}}
\caption{Very sick patient}\label{S5}
\end{figure*}

\begin{figure*}[!ht]
\centering%
\subfigure[\footnotesize Control $u$ (blue $-$) and Nominal control $u^\ast$ (black $- -$) ]
{\includegraphics[width=0.45\textwidth]{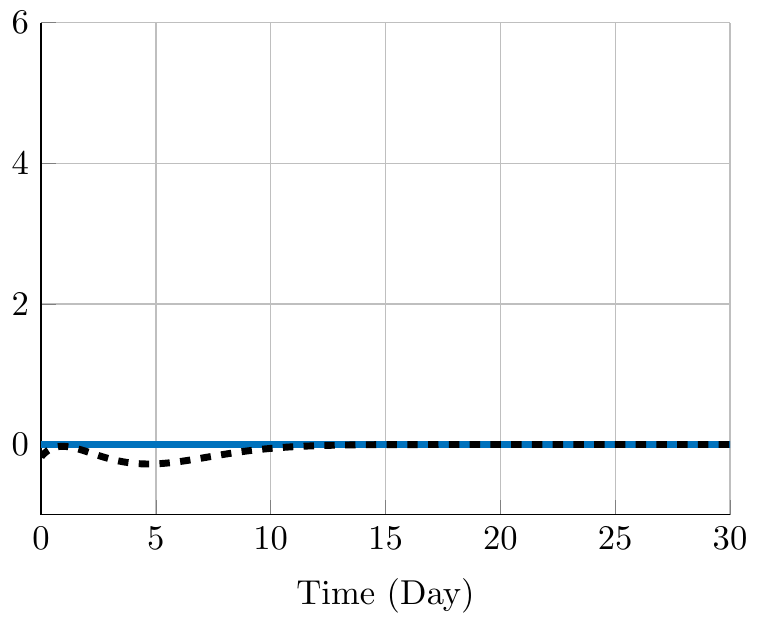}}
\subfigure[\footnotesize Control $v$ (blue $-$) and Nominal control $v^\ast$ (black $- -$)]
{\includegraphics[width=0.45\textwidth]{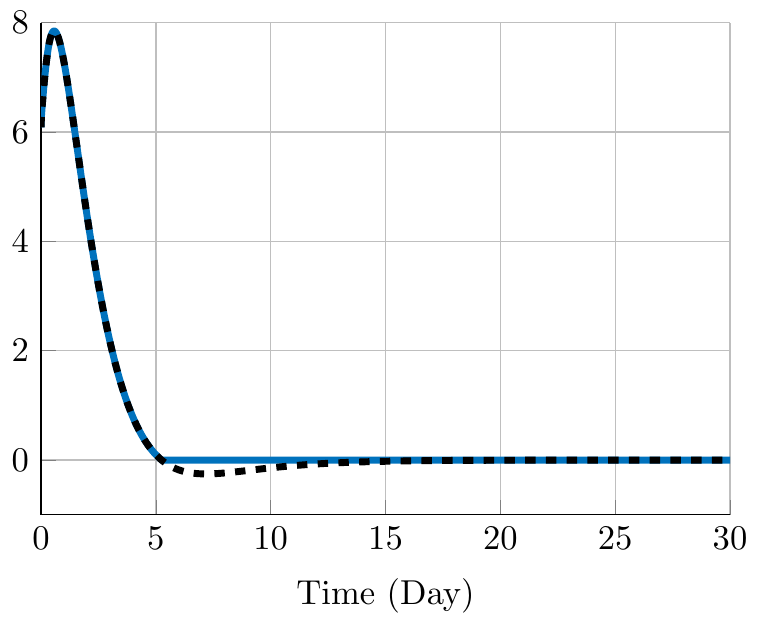}}
\\
\subfigure[\footnotesize Output $x$ (blue $-$), Reference trajectories (black $- -$) and Stable points (red and green $-.$)]
{\includegraphics[width=0.45\textwidth]{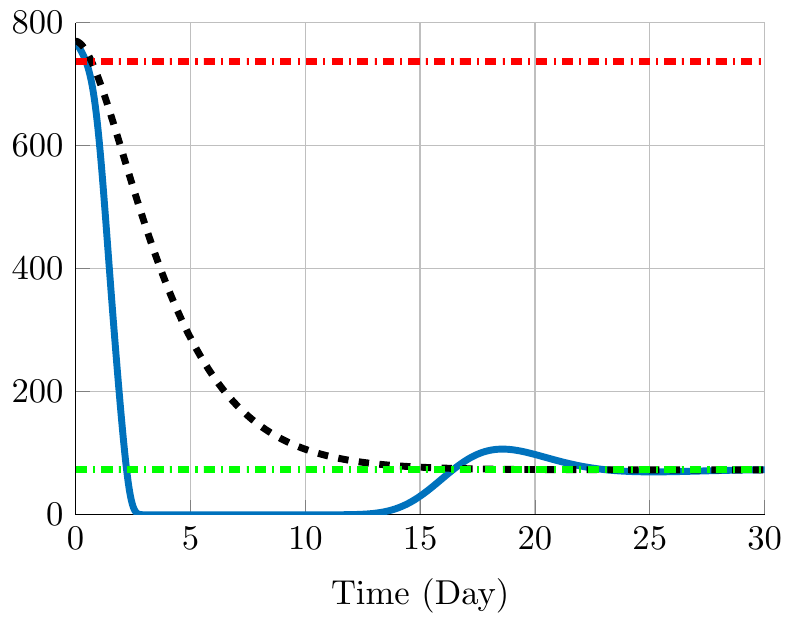}}
\subfigure[\footnotesize Output $y$ (blue $-$), Reference trajectories (black $- -$) and Stable points (red and green $-.$)]
{\includegraphics[width=0.45\textwidth]{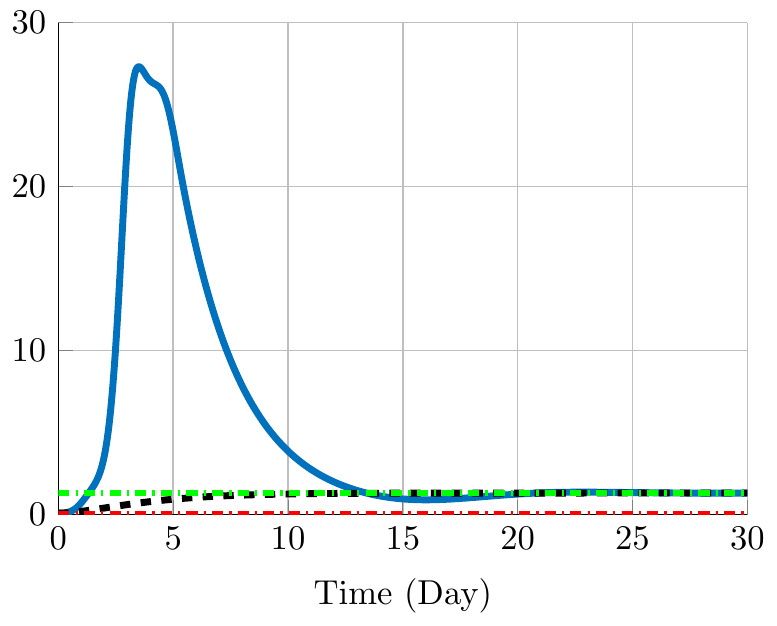}}
\caption{Open loop}\label{S6}
\end{figure*}

\begin{figure*}[!ht]
\centering%
\subfigure[\footnotesize Time evolution of $\eta_x$]
{\includegraphics[width=0.45\textwidth]{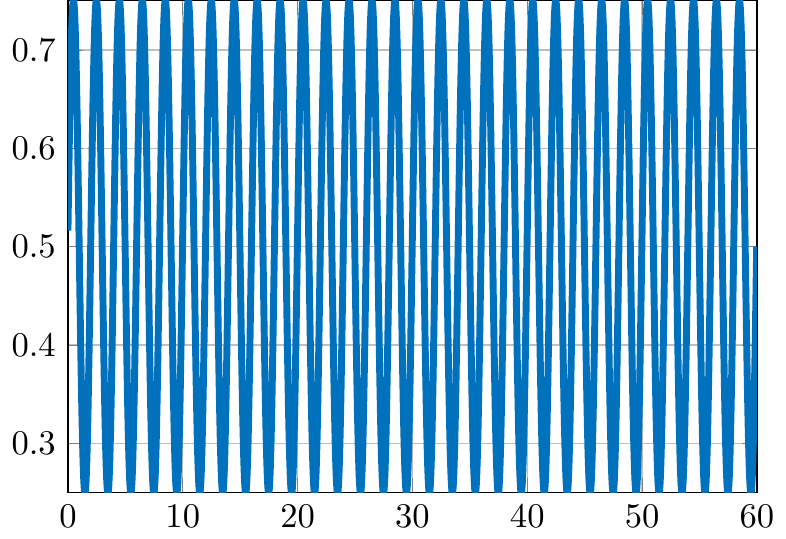}}
\subfigure[\footnotesize  Time evolution of $\eta_y$]
{\includegraphics[width=0.45\textwidth]{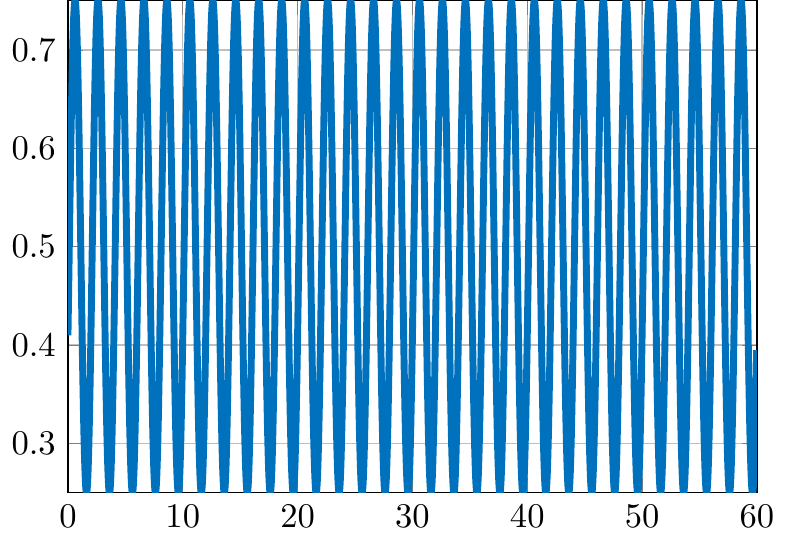}}
\caption{Fluctuation of the drug delivery}\label{S56}
\end{figure*}

The virtual patient is in a critical state, \textit{i.e.}, the initial state $x = 770$, $y = 0.1$ is close to the malignant equilibrium. The time variation of $\eta_x$ and 
$\eta_y$, which are displayed in Figure \ref{S56}, are assumed to be unknown. It is possible to cure the virtual patient without the cytotoxic drug, \textit{i.e.}, $u_{\rm CL} \equiv 0$.  Figure \ref{S5}, which should be of interest for concerologists, exhibits a convergence to the benign equilibrium with some oscillations perhaps due to the violent fluctuations of $\eta_y$. The quality of the open loop behavior in Figure \ref{S6} is lower.

\section{Conclusion}\label{conclusion}

\subsection{Main goal}
Some results encountered with our computer experiments might question oncologists:
\begin{itemize}
\item the quantity of injected drug might be lower in some slow scenario than in the corresponding fast one;
\item there are critical situations where only immunotherapy matters: the cytotoxic drugs are useless. 
 \end{itemize}
 Those startling calculations need of course to be further analyzed.
\subsection{Systems Biology}
In the spirit of \emph{Systems Biology} (see, \textit{e.g.}, \cite{vec}), let us suggest the the following research tracks:
\begin{itemize}
\item Examine parameter identification in Equation \eqref{1}.
\item Flatness-based control might be helpful elsewhere: 1) Another model due to \cite{hahn} has also been investigated from a control-theoretic perspective (see, \textit{e.g.}, \cite{kovacs14}, \cite{scha} and references therein, \cite{lin}).  It is easy to check that it is flat; 2) the unicycle in \cite{sharifi3}, which is used as a nanorobot for drug delivery, is well known to be flat. 
\subsection{New control paradigms?}
The control strategy which has been developed here for oncological in silico experiments might lead to new paradigms:
\begin{enumerate}
\item Assume that we have a flat nominal system with some important uncertainties. Use open-loop  flatness-based techniques. Close the loop via model-free control. What's about flatness-based control of partial differential equations (see, \textit{e.g.}, \cite{pde,meurer}; and references therein)?
\item Investigate possible connections with \emph{Active Disturbance Rejection Control}, or \emph{ADRC}, as presented by \newline \cite{adrc}.
\end{enumerate}


\end{itemize}



\newpage

\end{document}